\documentclass{ws-ijmpa}

\usepackage[super]{cite}
\usepackage{xcolor}
\usepackage[verbose,hypertexnames=false]{hyperref}
\hypersetup{colorlinks=false,allbordercolors=blue,pdfborderstyle={/S/U/W 1}}

\begin{document}

\markboth{Vindhyawasini Prasad}{Instructions for typing manuscripts (paper's title)}

%
\catchline{}{}{}{}{}
%

\title{Dark matter searches at BESIII}

\author{Vindhyawasini Prasad \\
(On behalf of the BESIII Collaboration)}

\address{Instituto de Alta Investigaci\'on \\
   Universidad de Tarapac\'a,\\
Casilla 7D, Arica 1000000, Chile \\
 vindymishra@gmail.com}

\maketitle

\begin{history}
\received{Day Month Year}
\revised{Day Month Year}
\accepted{Day Month Year}
\published{Day Month Year}
\end{history}

\begin{abstract}
Dark matter (DM) is a new type of invisible matter introduced to explain various features of recent astrophysical observations, including galaxy rotation curves and other fundamental characteristics of our universe. DM may couple to ordinary matter via portals, which open up possibilities for new particles, such as axion-like particle, light Higgs boson, dark photon, and spin-1/2 fermions. If the masses of these particles lie in the MeV to GeV range, they can be explored by high-intensity electron-positron collider experiments, such as the BESIII experiment. BESIII has accumulated a huge amount of datasets at several energy points, including the $J/\psi$, $\psi(3686)$, and $\psi(3770)$ resonances. These data samples have been utilized to explore the possibility for axion-like particle and light Higgs boson through radiative $J/\psi$ decays, dark photon via the initial-state radiation process, and a massless dark photon in $\Lambda_c$ decays. This report highlights the latest results from the BESIII experiment on these topics.

\end{abstract}


\ccode{PACS numbers: 03.65.$-$w, 04.62.+v}

\section{Introduction}	
The Standard Model (SM) is one of the most successful theories in describing the subatomic world at the fundamental level\cite{SM}. The discovery of the Higgs boson completed the last missing piece of the SM\cite{Higgs}. However, the SM is not a theory of everything because it is unable to explain the nature of recent experimental anomalies\cite{anomal1,anomal2}, including the lack of a viable dark matter (DM) candidate\cite{anomal1}. DM refers to a new kind of invisible matter that neither emits nor absorbs electromagnetic radiation. It accounts for about one-quarter of the universe's total matter density. However, its presence can only be inferred through its gravitational effects. Therefore, the nature of DM remains mysterious. A GeV-scale DM particle carying baryon number offers an attractive framework for understanding both the origin of DM and matter-antimatter asymmetry in the universe\cite{baryonDM}. Additionally, a massless dark photon\cite{MDM1,MDM2} could enhance the decay rate of flavor-changing neutral current (FCNC) processes, which are highly suppressed in the charm sector\cite{FCNC}. Therefore, an experimental evidence on DM is crucial for explaining the observed rotation curves of galaxies as well as astrophysical observations\cite{anomal1} and the formation of our universe.   

We do not yet fully understand how DM couples to SM particles. Many models beyond the SM propose a dark hidden sector\cite{Hidden}, where DM could couple to SM particles through various portals\cite{portal}. The possible  gauge bosons in this hidden sector include a light Higgs boson ($A^0$), an axion-like particle (ALP), a dark photon ($\gamma'$), or spin-1/2 sterile neutrinos. These particles are expected to have masses in the range of a few GeV, making them accessible via high-intensity $e^+e^-$ collider experiments, such as BESIII experiment\cite{bes3}. BESIII\cite{bes3}, a symmetric $e^+e^-$ collider, has collected a huge amount of datasets at several energy points from $2.0$ to $4.95$ GeV, including the $J/\psi$ and $\psi(3686)$ resonances. Utilizing these extensive datasets, BESIII has recently explored various DM scenarios\cite{vindy}, including searches for an axion-like particle\cite{alp1,alp2}, a light Higgs boson\cite{higgs1,higgs2} and a dark photon\cite{DP1,DP2}. This report reviews recent results from the BESIII experiment related to the exploration of these DM scenarios.

\begin{figure}[!pb]
  \centerline{\includegraphics[width=6.0cm,height=4.5cm]{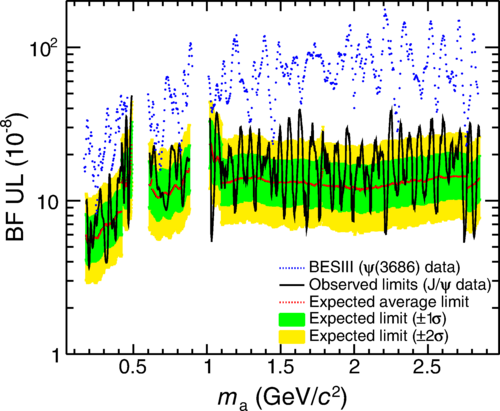}
\includegraphics[width=6.5cm,height=4.5cm]{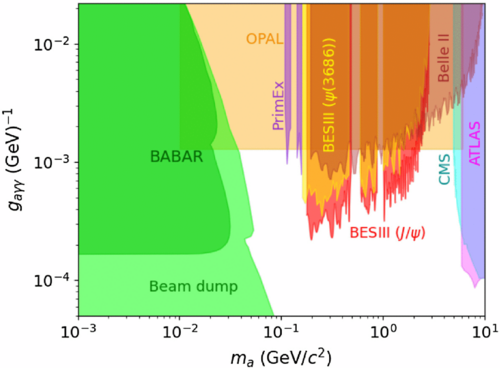}
  }

\vspace*{8pt}
\caption{The $95\%$ CL upper limits on product branching fractions $\mathcal{B}(J/\psi \to \gamma a) \times \mathcal{B}(a \to \gamma \gamma)$ (left) and axion-photon coupling $g_{a \gamma \gamma}$ (right) versus $m_a$. Our limits on $g_{a \gamma \gamma}$ are most stringent to date in the search mass region.}
\label{alp}
\end{figure}

\section{Search for an axion-like particle}
An axion was originally introduced via the spontaneous breaking of Peccei-Quinn symmetry\cite{Peccei} to address the strong $CP$\cite{strongCP} and hierarchy\cite{Graham} problems in quantum chromodynamics. Many extensions of the SM, such as an extended Higgs sector\cite{Branco} and string theory\cite{Ringwald}, predict the existence of ALPs. These ALPs ($a$) have the same spin-parity as the axion but differ in mass and coupling characteristics. ALPs primarily couple to a photon pair with a coupling constant $g_{a \gamma \gamma}$. At BESIII, the measurement of $g_{a \gamma \gamma}$ may procced via the ALP-strahlung process $e^+e^- \to \gamma a$\cite{Merlo} and radiative $J/\psi \to \gamma a$ decays\cite{Merlo,Masso}.

The current best upper bounds on $g_{a \gamma \gamma}$ have been obtained from the previous BESIII experiment\cite{alp1} for the mass range  $0.165 \le m_a \le 1.468$ GeV/$c^2$ and the OPAL experiment\cite{opal} for $1.468 \le m_a \le 5.0$ GeV/$c^2$. The previous BESIII measurement\cite{alp1} was based on approximately one billion $J/\psi$ events, which have been extracted from $2.7$ billion $\psi(3686)$ events by tagging the pion pair from the $\psi(3686) \to \pi^+\pi^- J/\psi$ transition. Recently, BESIII has performed a search for di-photon decays of an ALP via radiative $J/\psi \to \gamma a$ decays\cite{alp2}, using a dataset of 10 billion $J/\psi$ events. In addition to $J/\psi \to \gamma a$, this dataset may also include a $4.4\%$ contribution from the ALP-strahlung process $e^+e^- \to \gamma a$\cite{Merlo}, alongside dominant SM background contributions from $J/\psi \to \gamma P$ (where $P = \pi^0, \eta, \eta'$)\cite{pdg}.

The search for $a$ is conducted by performing a series of maximum likelihood fits to the di-photon invariant mass spectrum, $m_{\gamma \gamma}$, which includes all combinations of two-photon pairs after background rejection from $J/\psi \to \gamma P$. No evidence of ALP production is found, and we set $95\%$ confidence level (CL) upper limits on the product branching fractions $\mathcal{B}(J/\psi \to \gamma a) \times \mathcal{B}(a \to \gamma \gamma)$ and $g_{a \gamma \gamma}$ to be less than $(3.7 -48.5) \times 10^{-8}$ and $(2.2 -101.8) \times 10^{-4}$ GeV$^{-1}$, respectively, over the mass range  $0.18 \le m_a \le 2.85$ GeV/$c^2$, as shown in Fig.~\ref{alp}, after subtracting the ALP-strahlung $e^+e^- \to \gamma a$ contribution\cite{Merlo}. These limits are the most stringent to date within this mass region\cite{alp2}.

\begin{figure}[!pb]
  \centerline{\includegraphics[width=13.0cm,height=4.0cm]{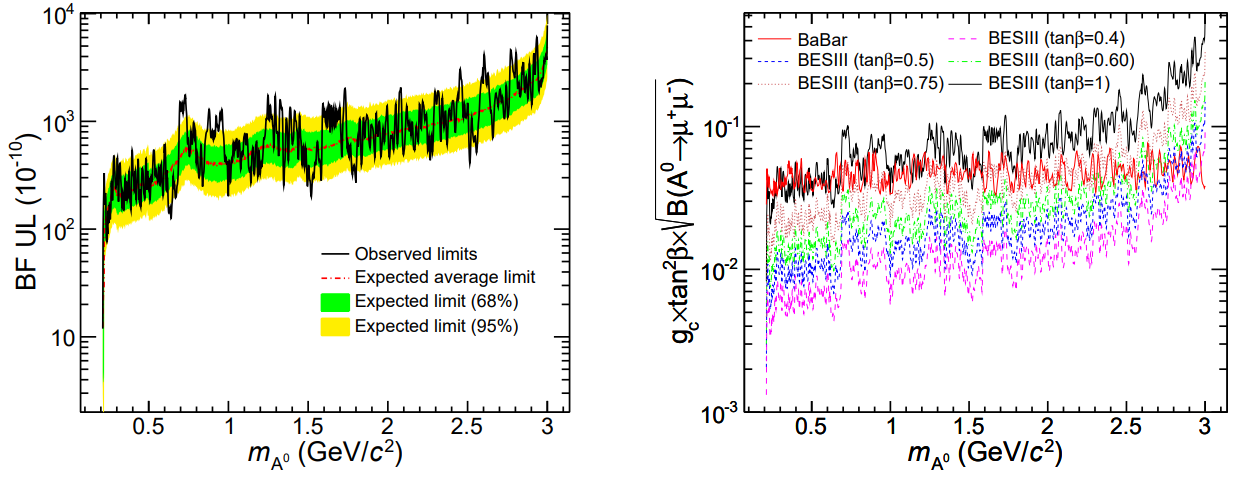} }

\vspace*{8pt}
\caption{The $90\%$ CL upper limits on product branching fractions $\mathcal{B}(J/\psi \to \gamma A^0) \times \mathcal{B}(A^0 \to \mu^+\mu^-)$ (left) and effective Yukawa coupling of the Higgs field to the bottom quark-pair (right). \label{higgs}}
\end{figure}

\section{Search for a CP-odd light Higgs boson}
A light Higgs boson can be a scalar or pseudoscalar particle predicted by various models beyond the SM, such as the Next-to-Minimal Supersymmetric Standard Model (NMSSM)\cite{Maniatis}. The NMSSM\cite{Maniatis} has a rich Higgs sector containing seven Higgs bosons, including a $CP$-odd light Higgs boson ($A^0$). The mass of $A^0$ is expected to be of the order of a few GeV, making it accessible via radiative $J/\psi$ decays\cite{Wilczek}. The coupling of $A^0$ to up- (down-) type quark pairs is proportional to $\cot\beta$ ($\tan \beta$), where $\tan\beta$ represents the ratio of the vacuum expectation values of the up- and down-type Higgs doublets. The branching fraction of $J/\psi \to \gamma A^0$ is predicted to be within the range of $10^{-9}-10^{-7}$, depending on $m_{A^0}$, $\tan \beta$, and other NMSSM parameters\cite{Dermisek}.

Searches for $A^0$ have been explored in various decay channels by several collider experiments, including BaBar\cite{BaBar} and BESIII\cite{higgs1}, but so far only negative results have been reported. BESIII recently conducted a search for $A^0$ via radiative $J/\psi$ decays using a dataset of 9 billion $J/\psi$ events\cite{higgs2}. No significant signal for $A^0$ production is found, and we set $90\%$ CL upper limit on the product branching fractions $\mathcal{B}(J/\psi \to \gamma A^0) \times \mathcal{B}(A^0 \to \mu^+\mu^-)$ in the range of $(1.2-778.0)\times10^{-9}$ for $0.212 \le m_A \le 3.0$ GeV/$c^2$, as shown in Fig.~\ref{higgs} (left). This new BESIII result\cite{higgs2} improves upon the previous BESIII measurement\cite{higgs1} by a factor of $6-7$. It also surpasses the BaBar measurement\cite{BaBar} in the low-mass region for $\tan \beta =1$, as shown in the plot of the $90\%$ CL upper limits on the effective Yukawa coupling of the Higgs field to bottom quark pairs versus $m_{A^0}$\cite{higgs2} (Fig.~\ref{higgs} (right)).

\begin{figure}[!pb]
  \centerline{\includegraphics[width=6.0cm,height=4.5cm]{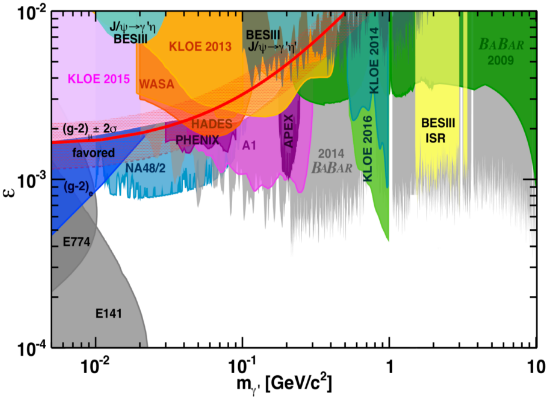}
\includegraphics[width=6.5cm,height=4.5cm]{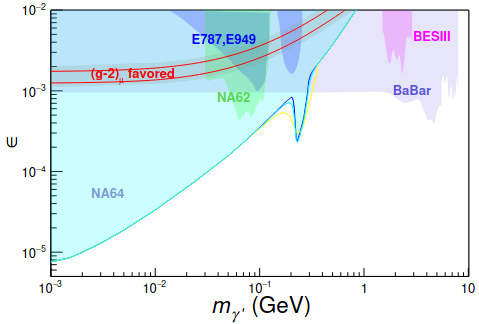}
  }

\vspace*{8pt}
\caption{The $90\%$ CL upper limits on coupling strength between dark photon and SM particles ($\epsilon$) for dark photon decaying to visible final lepton-pair particles (left) and invisible final state particles (right). An untagged photon method technique has enabled better exclusion limits on $\epsilon$ in di-lepton decays of dark photon in the ISR process of $e^+e^- \to \gamma_{\rm ISR} \gamma'$. \label{DP}}
\end{figure}

\section{Sarch for dark photon}
The hidden dark sector model predicts a new type of force carrier called the dark photon ($\gamma'$) in the simplest scenario of an Abelian $U(1)$ gauge field\cite{Hidden}. The dark photon may couple to SM particles through a kinetic mixing parameter, defined as $\epsilon = \alpha'/\alpha$, where $\alpha$ and $\alpha'$ are the fine structure constants of the SM and dark sectors, respectively. The mass of the dark photon is expected to be a few GeV to satisfy both astrophysical constraints\cite{anomal1} and the observed deviation in the muon’s anomalous magnetic moment\cite{amu}.

Searches for the $\gamma'$ have been conducted by various experiments for both its visible and invisible decay modes\cite{vindy}. However, so far, only null results have been reported, with $90\%$ CL exclusion limits on $\epsilon$ set down to the $10^{-3}$ level\cite{vindy}. Some of the most stringent exclusion limits on $\epsilon$ for both visible and invisible decay modes of $\gamma'$ in the mass range between approximately 100 MeV/$c^2$ and 10 GeV/$c^2$ come from the BaBar experiment\cite{babar1,babar2}.

Recently, BESIII has conducted a search for di-lepton decays of the dark photon via initial state radiation (ISR) process $e^+e^- \to \gamma_{\rm ISR} \gamma'$, $\gamma' \to e^+e^-, \mu^+\mu^-$, using a data sample of 2.93 fb$^{-1}$ collected at the $\psi(3770)$ resonance\cite{DP1}. An untagged photon technique was employed, which excludes the ISR photon from the detector acceptance region. No evidence of $\gamma'$ production is found, and we set $90\%$ CL upper limit on $\epsilon$ at the level of $10^{-3}$, as shown in Fig.~\ref{DP} (left). This untagged photon technique allowed BESIII to achieve limits comparable to those of the BaBar experiment\cite{babar1}, which utilized a data sample of 514 fb$^{-1}$ collected at the $\Upsilon$ resonance.

More recently, BESIII has also explored the possibility of invisible decays of the dark photon via the ISR process $e^+e^- \to \gamma_{\rm ISR} \gamma'$, using 14.9 fb$^{-1}$ of $e^+e^-$ annihilation data collected at center-of-mass (CM) energies from 4.13 to 4.6 GeV\cite{DP2}. No evidence of invisible $\gamma'$ decays is found, and we set $90\%$ CL upper limit on $\epsilon$, which is slightly above the existing BaBar measurement, as shown in Fig. 2. With the recently collected 20 fb$^{-1}$ data sample at the $\psi(3770)$ resonance\cite{bes3}, we expect to achieve results that may surpass the BaBar sensitivity\cite{babar2}.

\begin{figure}[!pb]
  \centerline{\includegraphics[width=12.0cm,height=4.0cm]{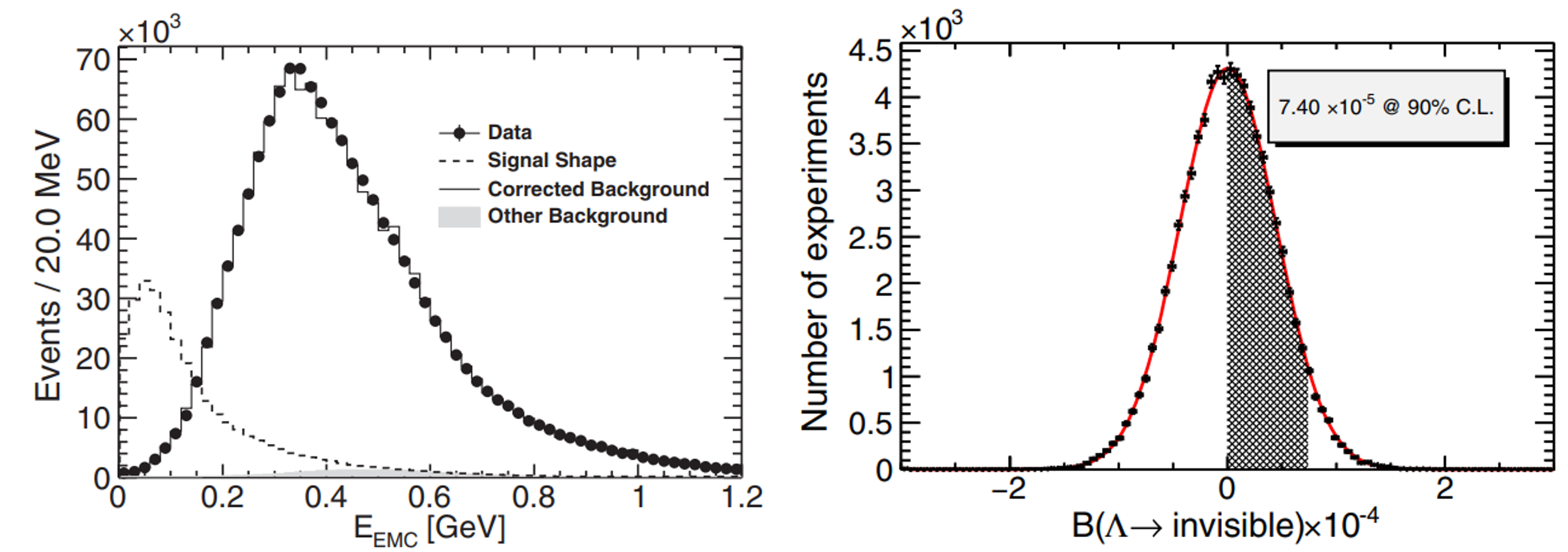}
  }

\vspace*{8pt}
\caption{(Left) The total energy deposited in the EMC ($E_{\rm EMC}$) (black dots with error bars), signal MC (dashed curve) and background MC from $\Lambda \to n \pi^0$ decay (solid line). The right hand side plot shows the distribution of the estimated branching fraction $\mathcal{B}(\Lambda \to {\rm invisible})$ obstained from pseudosamples. \label{lambdabaryon}}
\end{figure}

\section{Invisible decays of \boldmath{$\Lambda$} baryon}
 Baryon number violation is considered to be a major cause of the observed matter-antimatter asymmetry in the universe. The DM density is expected to be approximately 5.4 times the baryon matter density. Therefore, invisible baryon decays, such as $\Lambda \to {\rm invisible}$\cite{baryonDM}, may be contributed by the DM.

BESIII has recently performed the  search for invisible decays of the $\Lambda$ baryon using a dataset of 10 billion $J/\psi$ events in the process $J/\psi \to \Lambda \bar{\Lambda}$ for the first time\cite{lambdabaryon}. This study employed a double-tag (DT) technique. One of the $\Lambda$ baryons is tagged by its dominant hadronic decay mode, $\Lambda \to p \pi$, while the recoil side has been used to infer invisible decays of the $\Lambda$ baryon. Signal events for invisible decays are extracted by fitting the total energy deposited in the electromagnetic calorimeter (EMC), denoted as $E_{\rm EMC}$.

Because of dominant contribution of $\Lambda \to n \pi^0$, the $E_{\rm EMC}$ distribution consists of three components: $E_{\rm EMC} = E_{\rm EMC}^{\pi^0} + E_{\rm EMC}^n + E_{\rm EMC}^{\rm noise}$, where $E_{\rm EMC}^{\pi^0}$, $E_{\rm EMC}^n$, and $E_{\rm EMC}^{\rm noise}$ represent the energy deposited in the EMC from electromagnetic showers due to $\pi^0$ decays, neutrons, and background noise, respectively. For the invisible decay signal, the $E_{\rm EMC}$ distribution peaks at zero, while it deviates from zero for the backgrounds from $\Lambda \to n \pi^0$ decays, as shown in Fig. 3 (left).

The fit yields no evidence for invisible decays of $\Lambda$ baryon, and therefore we set the $90\%$ CL upper limit on the branching fraction of $\Lambda$ baryon $\mathcal{B}(\Lambda \to {\rm invisible})$ to be less than $7.4 \times 10^{-5}$, as seen in Fig. 3 (right).

\begin{figure}[!pb]
  \centerline{\includegraphics[width=12.0cm,height=4.0cm]{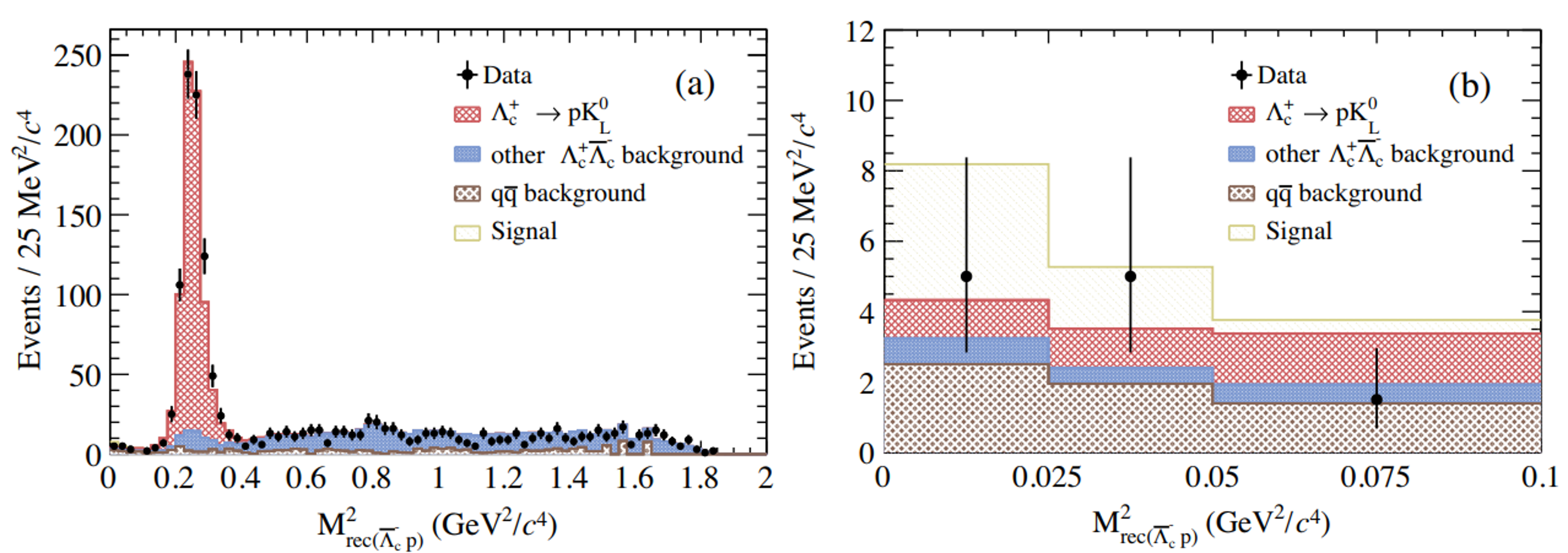}
  }

\vspace*{8pt}
\caption{The distribution of $M^2_{\rm rec(\bar{\Lambda}_c^-p)}$ of the accepted DT candidate events for data (black dots with error bars), signal MC (gray) and various other background predictions shown by pattern colored histographs in full spectrum (left) and signal region (right). \label{masslessDP}}
\end{figure}

\section{Search for massless dark photon}
Many models beyond the SM predict a massless dark photon ($\gamma'$) introduced by the spontaneous breaking of an Abelian $U(1)_D$ group\cite{MDM1,MDM2}. Such a massless dark photon may contribute to FCNC processes in the charm sector to potentially enhance their decay rates\cite{FCNC}. This massless dark photon can be explored through the two-body decay $\Lambda_c \to p \gamma'$.

Recently, a search for the massless dark photon has conducted via the process $e^+e^- \to \Lambda_c^+\bar{\Lambda}_c^-$, with $\Lambda_c^+ \to p \gamma'$, using 4.5 fb$^{-1}$ of data collected at CM energies between 4.6 and 4.699 GeV by the BESIII detector\cite{masslessDP}. A DT technique is used for this search, where one of the $\Lambda_c$ baryons is reconstructed using its dominant hadronic decay modes, while the other $\Lambda_c$ is reconstructed from the decay $\Lambda_c^+ \to p \gamma'$.

In DT events, the square of the recoil mass, $M^2_{\rm rec}(\bar{\Lambda}_c^-p)$, against the single-tagged $\bar{\Lambda}_c^-$ and the proton $p$, is used to infer the massless dark photon signal. No significant signal events for the massless dark photon are observed, and we set the  $90\%$ CL upper limit on $\mathcal{B}(\Lambda_c^+ \to p \gamma')$ to be less than $8 \times 10^{-5}$ for the first time.

\section{Summary}
BESIII has recently conducted a series of searches for various flavors of dark matter particles using data samples collected at several energy points in the tau-charm region. No significant signal events are observed for these scenarios, and we set one of the most stringent stringent exclusion limits. These limits exclude a large fraction of the parameter space for new physics models beyond the SM. Many new results on this topic are anticipated in the near future, especially with the recently collected 20 fb$^{-1}$ of $\psi(3770)$ data.  

\section*{Acknowledgments}
Vindhyawasini Prasad acknowledges the partial financial support received from the Agencia Nacional de Investigación y Desarrollo (ANID), Chile, from ANID FONDECYT regular 1230987 Etapa 2023, Chile, and from ANID PIA/APOYO AFB220004, Chile.

\section*{ORCID}

\noindent Vindhyawasini Prasad - \url{https://orcid.org/0000-0001-7395-2318}

\end{document}